\documentstyle[aps,pra,preprint,epsfig,tighten]{revtex}

\draft

\def\bra#1{{\langle#1|}}
\def\ket#1{{|#1\rangle}}
\def\expect#1{{\langle#1\rangle}}

\def\w{{\vec w}}

\def\tr{{\rm Tr}}
\def\psim{{\bf\Psi_-}}
\def\psip{{\bf\Psi_+}}
\def\psipm{{\bf\Psi_\pm}}
\def\phim{{\bf\Phi_-}}
\def\phip{{\bf\Phi_+}}
\def\phipm{{\bf\Phi_\pm}}
\def\psucc{{p_s}}
\def\Nsucc{{N_s}}
\def\CK{{\rm CK}}
\def\ovec{{\vec{\,o}\,}}
\def\ivec{{\vec{\,\imath}\,}}
\def\j{{\vec{\,\jmath}\,}}

\begin{document}

\preprint{IASSNS-HEP-00/79}

\title{Entanglement purification of unknown quantum states}

\author{Todd A. Brun,$^{1,}$\thanks{Current address:  Institute for Advanced
Study, Einstein Drive, Princeton, NJ 08540.}
Carlton M. Caves,$^2$ and R\"udiger Schack$^3$}

\address{
$^1$Physics Department, Carnegie Mellon University, Pittsburgh, PA 15213, USA\\
$^2$Department of Physics and Astronomy, University of New Mexico,\\
Albuquerque, New Mexico 87131-1156, USA \\
$^3$Department of Mathematics, Royal Holloway, University of
    London, Egham, Surrey TW20 0EX, UK}

\date{9 October 2000}

\maketitle

\begin{abstract}
A concern has been expressed that ``the Jaynes principle can
produce fake entanglement'' [R.~Horodecki {\it et al.}, Phys.\
Rev.\ A {\bf 59}, 1799 (1999)].  In this paper we discuss the
general problem of distilling maximally entangled states from $N$
copies of a bipartite quantum system about which only partial
information is known, for instance in the form of a given
expectation value. We point out that there is indeed a problem
with applying the Jaynes principle of maximum entropy to more than
one copy of a system, but the nature of this problem is classical
and was discussed extensively by Jaynes. Under the additional
assumption that the state $\rho^{(N)}$ of the $N$ copies of the
quantum system is {\it exchangeable}, one can write down a simple
general expression for $\rho^{(N)}$. We show how to modify two
standard entanglement purification protocols, one-way hashing and
recurrence, so that they can be applied to exchangeable states. We
thus give an explicit algorithm for distilling entanglement from
an unknown or partially known quantum state.
\end{abstract}


\section{Introduction}

Entanglement is a quantum-mechanical resource that can be used for a
number of tasks, including quantum teleportation, quantum cryptography,
and quantum dense coding. Since real quantum channels are noisy, it is
very difficult to create perfect entanglement directly between two
distant parties.  There is thus a need to purify (or distill) partial
entanglement.  Suppose two parties share $N$ pairs of qubits
such that each pair is in the same entangled, but mixed state $\rho$,
the total state $\rho^{(N)}$ of all $N$ pairs thus being the $N$-fold
tensor product $\rho^{(N)}=\rho^{\otimes
N}\equiv\rho\otimes\ldots\otimes\rho$. There exist protocols
\cite{Bennett1996b,Bennett1996c,Horodecki1997b,Deutsch1998},
using only local operations
and classical communication, which allow the two parties to transform
$M<N$ of the pairs into maximally entangled states, for instance
singlet states. In the limit $N\to\infty$, the fidelity of the
singlets approaches 1 and the fraction $M/N$ a fixed limit, called the
asymptotic yield.

In this paper, we consider the more general case in which the initial state
$\rho^{(N)}$ is not a tensor-product state. This corresponds to the realistic
situation that the state $\rho$ of each individual pair is not perfectly
known, for instance because one of the particles has been sent through a
channel with only partially known characteristics.  In Secs.~\ref{sec:hashing}
and \ref{sec:recurrence}, we apply the entanglement
purification methods known as one-way hashing \cite{Bennett1996c} and
recurrence \cite{Bennett1996c,Deutsch1998} to partially known, including
completely unknown, quantum states. It turns out that the generalization of
the recurrence method is straightforward, whereas the hashing method as it is
described in Ref.~\cite{Bennett1996c} depends on the initial state being of
tensor-product form and therefore requires a more careful analysis.  Unlike
Briegel {\it et al.}\ \cite{Giedke1999}, who have studied entanglement
purification with imperfect quantum operations, we assume that all operations
are error-free. A paper related to ours is Ref.~\cite{Eisert2000a} by Eisert
{\it et al.}, who study how distillable entanglement decreases
when information about a quantum state is lost.

Before we turn to the actual entanglement purification protocols, we discuss,
in Sec.~\ref{sec:assign}, the problem of what density operator $\rho^{(N)}$ to
assign to $N$ pairs of qubits if only partial information is available. This
is an unsolved problem, and we do not attempt to give a general solution. We
show, however, that under the additional assumption of {\it exchangeability\/}
the state $\rho^{(N)}$ must have a certain simple form,
which is amenable to entanglement purification.  Our discussion also
provides a resolution of the apparent paradox found by Horodecki {\it et
  al.}\ \cite{Horodecki1999b}, who give an example where applying the
Jaynes principle of maximum entropy \cite{Jaynes1957a,Jaynes1957b} leads to a
state with more distillable entanglement than seems to be warranted by the
available information.  We conclude in Sec.~\ref{sec:conclude}.

\section{State assignment based on partial information}  \label{sec:assign}

Let us consider the example given by Horodecki {\it et al.}\
\cite{Horodecki1999b}.  The authors consider a system composed of a single
pair of qubits and define an operator
\begin{equation}
B = {1\over2}(\sigma_x\otimes\sigma_x + \sigma_z\otimes\sigma_z)
  = ( \phip - \psim ) \;,
\end{equation}
where
$\psipm=\ket{\Psi_\pm}\bra{\Psi_\pm}$, $\phipm=\ket{\Phi_\pm}\bra{\Phi_\pm}$
are projectors onto the Bell states,
\begin{eqnarray}
\ket{\Phi_\pm} &=& {1\over\sqrt2}(\ket{00}
  \pm \ket{11})\;, \nonumber\\
\ket{\Psi_\pm} &=& {1\over\sqrt2}(\ket{01}
  \pm \ket{10})\;.
\label{Bellbasis}
\end{eqnarray}
Our definition of $B$ differs from that of Ref.~\cite{Horodecki1999b} by a
constant factor to simplify the expressions.
If {\it all that is known\/} about the system state is the expectation
value $\expect B =1/2$, then Jaynes's principle of maximum entropy
stipulates that one should assign the state of maximum von Neumann
entropy compatible with the constraint  $\expect B=1/2$, which in this
case is
\begin{equation}
\rho_J = {9\over16}\phip+{1\over16}\psim + {3\over16}( \psip + \phim ) \;.
\end{equation}
This state has distillable entanglement. Horodecki {\it et al.}\
\cite{Horodecki1999b} point out that the state
\begin{equation}
\rho_H = {1\over2}\phip+{1\over4}( \psip + \phim ) \;.
\end{equation}
also satisfies the constraint $\expect B=1/2$, but is separable and hence
unentangled.  They conclude that the
entanglement in the maximum entropy state $\rho_J$ is ``fake,'' because it
violates the condition that an inference scheme ``should not give us an
inseparable estimated state if only theoretically there exists a separable state
consistent with the measured data.'' As an alternative to the Jaynes principle,
they propose first to minimize the entanglement and then to find the state of maximum
entropy among those states that have minimal entanglement. For the constraint
$\expect B=1/2$, this alternative scheme results in the state $\rho_H$ given
above.

A simple defense of the Jaynes principle would be the following (see
also Refs.\ \cite{Rajagopal1999,Rigo2000}). The alternative procedure
proposed by Horodecki {\it et al.}\ assumes additional information about the
two qubits, namely that entanglement is {\it a priori\/}
unlikely. This would be reasonable, e.g., in a situation where the
parties know that the state has been prepared by an adversary whose
objective is to let them have as little entanglement as possible. But
then more is known about the state than just the given expectation
value, and hence the assumptions behind the Jaynes procedure are not
fulfilled.

If there is no specific additional information, however, the maximum
entropy state assignment $\rho_J$ is preferable to the minimum
entanglement assignment $\rho_H$. Indeed, if a projective measurement
in the Bell basis is performed, assigning $\rho_H$ corresponds to
assigning zero probability to the measurement outcome $\psim$, an outcome
that is not ruled out by the constraint $\expect B =1/2$. In this sense, the
minimum entanglement assignment is inconsistent with the prior information.
By contrast, no inconsistency of this kind can arise from
the maximum entropy assignment in the absence of prior information
beyond the given expectation value.
Since no measurement of a single system can tell if that system is entangled
or not, the prediction of ``fake entanglement'' for $\rho_J$ can cause no
difficulty.  In particular, there is no way to turn a single $\rho_J$ state
into a maximally entangled state even probabilistically \cite{Linden1998b}.

We now turn to the case in which the parties share not just one, but $N$ qubit
pairs. We denote by $\rho^{(N)}$ the total state of the $N$
pairs and assume that the $N$ pairs are known to satisfy the
constraints $\langle B\rangle_k\equiv\tr (\rho_k B)=1/2$ for
$k=1,\ldots,N$, where $\rho_k$ is the reduced density operator of the $k$-th
pair.  In this case, the state assignment $\rho^{(N)}=\rho_J^{\otimes N}$ is
not supported by the prior information, even though this is the state
of maximum entropy compatible with the given expectation values. For
large $N$, this state assignment corresponds to the definite
prediction that a nonzero number of perfect singlets can be distilled,
which is certainly not implied by the given expectation values.
The alternative state assignment $\rho^{(N)}=\rho_H^{\otimes
N}$ would, however, be equally unsupported by the prior information.
It corresponds to the definite prediction that {\it no\/} singlets can be
distilled from the $N$ pairs, which
is the minimum number of distillable singlets compatible with the {\it
a priori\/} knowledge. Although this is a very cautious prediction, it
is also not implied by the given expectation values.

The fact that a na{\"\i}ve application of the principle of maximum entropy
to many copies of a system fails is essentially of classical origin
and is not unique to problems involving entanglement. Jaynes
\cite{Jaynes1986} has given a thorough discussion of this problem,
which can be explained by a simple example. Consider a possibly loaded
die. All that is known about the die is the mean value $\langle
n\rangle\equiv\sum_n np(n)=3.5$, where $p(n)$ is the
probability of the outcome $n$, $n=1,\ldots,6$. The probability
distribution of maximum entropy compatible with the given mean-value
constraint is $p(n)=1/6$ for $n=1,\ldots,6$.
Now consider throwing the die $N$ times. A na{\"\i}ve application of the
maximum entropy principle would predict that
the $N$ dice throws were independent and identically distributed
(i.i.d.) according to the single-trial distribution $p(n)$.  This would
lead to the prediction that the fraction of throws showing any particular
outcome would approximate 1/6 with arbitrary precision as $N$ tended to
infinity. This prediction, however, is not implied by the prior knowledge,
which is compatible with many possible outcome sequences, including sequences in
which only the events $n=1$ and $n=6$ ever occur---quite possible, if the die
is loaded.  Moreover, with an i.i.d.\ distribution, the results of earlier
throws imply nothing about the probability of later outcomes.  Even
the most gullible gambler might become suspicious if 1 and 6 were
the only outcomes after thousands of throws.

In Ref.\ \cite{Jaynes1986}, Jaynes discusses how to
choose the multi-trial distribution in the classical case.
The starting point of his discussion is the assumption that the probability
distribution of the $N$ dice throws is {\it exchangeable}. The same
assumption is the starting point for our quantum analysis.
If exchangeability is assumed, the task of assigning a state of
$N$ qubit pairs compatible with the constraints given above is much
simplified.  A state $\rho^{(N)}$ of $N$ copies of a system is exchangeable
if it is a member of an exchangeable sequence $\rho^{(k)}$,
$k=1,2,\ldots$~.  An exchangeable sequence is defined by
\begin{list}
{(i)}\item $\rho^{(k)}=\tr_{k+1}\rho^{(k+1)}$
for all $k$, where $\tr_{k+1}$
denotes the partial trace over the $(k+1)$th system, and
\end{list}
\begin{list}
{(ii)}\item each $\rho^{(k)}$ is invariant under permutations
of the $k$ systems on which it is defined.
\end{list}
\vspace{10pt}
\noindent This definition is the quantum generalization
of de~Finetti's \cite{DeFinetti1990} definition of exchangeable sequences of
classical random variables.

A state $\rho^{(N)}$ is exchangeable if and only if it can be written in the
form
\begin{equation}
\rho^{(N)} = \int d\rho\,p(\rho) \rho^{\otimes N} \;,
\label{eq:exch}
\end{equation}
where $d\rho$ is a measure on density operator space, and $p(\rho)$ is a
normalized generating function, $\int d\rho\, p(\rho)=1$. This is a
consequence of the quantum de Finetti theorem, the quantum version of the
fundamental representation theorem due to de~Finetti \cite{DeFinetti1990}. The
quantum theorem was first proved by Hudson and Moody \cite{Hudson1976} after
pioneering work by St{\o}rmer \cite{Stormer1969}; for an elementary
proof see Ref.~\cite{Caves2000a}.

How, in general, do we pick $p(\rho)d\rho$? To our knowledge, there is no
universal rule for this task, although there exist a number of proposals for
unbiased measures $d\rho$ on density operator space
\cite{Bures1969,Braunstein1994a,Slater1997,Zyczkowski1998b}.  These can be
interpreted as proposals for state assignments for $N$ systems under the sole
assumption of exchangeability, i.e., using a generating function $p(\rho)=1$.

If, in addition to exchangeability, there is  a mean-value constraint
$\expect O=o$, the na{\"\i}ve Jaynes maximum entropy state assignment leads
to a generating function of the form $p(\rho) = \delta(\rho-\rho_J)$, where
$\rho_J$ is the single-system state of maximum entropy, subject to the constraint;
this generating function is unacceptable for the reasons given above.  A good
choice of $p(\rho)d\rho$ should be nonzero for all $\rho$ that are compatible
with the prior information---we should never arbitrarily rule out any possibility.
Similarly, $p(\rho)d\rho$ ought to vanish for any $\rho$ which is actually
ruled out by the prior information.  We therefore would expect a multi-system
generalization of Jaynes's maximum entropy procedure to have the form
\begin{equation}
p_{\scriptscriptstyle{\rm MAXENT}}(\rho) d\rho
  = {\cal N} \delta[o - \tr(O\rho) ] f(\rho) d\rho,
\label{multi_maxent}
\end{equation}
where ${\cal N}$ is a normalization constant and $f(\rho)d\rho$ is strictly
positive.  The exact form of the function $f(\rho)$ and of the measure
$d\rho$ is the subject of ongoing research.
In the spirit of the single-system Jaynes principle, $p(\rho)d\rho$
should favor states $\rho$ with higher von Neumann entropy $S(\rho)$
and should give the usual $\rho_J$ when $N=1$ \cite{Skilling1989a}.

Given an initial state assignment of the form (\ref{eq:exch}), additional
information can be obtained, e.g., by making measurements on individual
subsystems. Suppose a measurement outcome $k$ is represented by a positive
single-system operator $F_k$, with $\sum_k F_k=1$; i.e., the $F_k$ form a
positive-operator valued measure (POVM) \cite{Kraus1983}.
Given that the subsystem is in state $\rho$, the
probability of getting outcome $k$ is  $p(k|\rho) = \tr[F_k\rho]$.  If
the total state is given by Eq.~(\ref{eq:exch}), the probability of outcome
$k$ in a measurement on a single subsystem is then
\begin{equation}
p_k = \int d\rho\, p(\rho) p(k|\rho) \;.
\end{equation}
After the measurement we must update the state of
the remaining $N-1$ systems by Bayes's rule,
\begin{equation}
\rho^{(N-1)} = \int d\rho\, p(\rho|k) \rho^{\otimes(N-1)}\;,
\end{equation}
where \cite{Schack2000a}
\begin{equation}
p(\rho|k)={p(\rho)p(k|\rho)\over p_k}\;.
\end{equation}
By doing different
measurements on several subsystems, we acquire more and more data; if these
measurements are chosen well, the resulting posterior $p_{\rm post}(\rho)$
becomes more and more peaked and has less and less
dependence on the choice of prior $p(\rho)$.  This procedure is a
straightforward Bayesian version of quantum state tomography
\cite{Vogel1989b,Smithey1993,Leonhardt1995}.

The condition of exchangeability in combination with the quantum
de~Finetti theorem provides only a partial solution of the problem of state
assignment in the presence of partial information, but we show in the next two
sections that exchangeability alone is sufficient to guarantee that the
entanglement purification procedures known as one-way hashing and recurrence
can be carried out. The probability of distilling a positive yield of maximally
entangled states depends on the exact form of $p(\rho) d\rho$ in
Eq.~(\ref{eq:exch}).

\section{Entanglement purification by one-way hashing} \label{sec:hashing}

In this section, we first present a version of the one-way hashing
algorithm that proceeds by Bayesian updating of the probabilities for
products of Bell states and that can in principle be applied to general
exchangeable states. We then briefly sketch the argument given in
Ref.~\cite{Bennett1996c} that for a product state $\rho^{\otimes N}$, where
$\rho$ is Bell-diagonal with von~Neumann entropy $S$, the asymptotic yield of
pure singlets is given by $N(1-S)$. We show how to modify this argument
so that it can be applied to general exchangeable states. Finally we give a
simplified Bayesian hashing algorithm for exchangeable states and discuss
its asymptotic yield.
Our analysis is restricted to pairs of qubits, but the method generalizes
straightforwardly to arbitrary Hilbert space dimensions.

We restrict attention to Bell-diagonal states, i.e., mixtures of the Bell
states,
\begin{equation}
\rho_\w = w_1 \psim + w_2 \psip + w_3 \phim + w_4 \phip\;,
\label{bell_mixtures}
\end{equation}
where we denote the weights by $\w=\{w_1,w_2,w_3,w_4\}$, $w_1+w_2+w_3+w_4=1$,
$w_j\ge0$ for $j=1,\ldots,4$.  Most existing entanglement purification
procedures begin by making this assumption.  If it does not hold, it is
possible to put any state in this form by ``twirling,'' that is, by randomly
rotating both spins of an entangled pair.  The final yield of maximally
entangled states cannot be diminished by omitting this step, however, so it
is better to think of twirling as a conceptual, rather than a physical
procedure.  After twirling, the initial, exchangeable state (\ref{eq:exch})
of our $N$ pairs of qubits becomes
\begin{equation}
\rho^{(N)} = \int d\w\, p(\w) \rho_\w^{\otimes N} \;,
\label{rhon}
\end{equation}
where
\begin{equation}
   \int d\w\, p(\w) = 1 \;.
\label{rhon_norm}
\end{equation}
We now define the set of labeled states
\begin{eqnarray}
\rho_{00} = \psim\;, &\ \ \ & \rho_{01} = \psip\;, \nonumber\\
\rho_{10} = \phim\;, &\ \ \ & \rho_{11} = \phip\;.
\end{eqnarray}
The first bit in the label tells us whether the pair is in a $\Psi$ or a
$\Phi$ state; the second bit tells us whether it is in a $+$ or $-$ state.  If
we are restricted to local measurements and classical communication on a
single pair, the best we can do is to determine one of these two bits, but not
both, and the pair will be left in an unentangled state.  Bennett {\it et al.}\
have shown, however, that if we can manipulate the qubits collectively, much
more interesting measurements are possible \cite{Bennett1996c}.

The first step is to rewrite the state (\ref{rhon}) as a probability
distribution over strings of bits, with each qubit pair associated
with two bits in the string.
For this, we define the product distribution
\begin{equation}
p(i_1i_2\cdots i_{2N}|\w) =  w_{i_1i_2} w_{i_3i_4} \cdots w_{i_{2N-1}i_{2N}}\;,
\end{equation}
where $w_{00} \equiv w_1$, $w_{01} \equiv w_2$, $w_{10} \equiv w_3$, and
$w_{11} \equiv w_4$. Using this notation,
\begin{equation}
\rho^{(N)} = \sum_{i_1,i_2,\ldots,i_{2N}} p(i_1i_2\ldots i_{2N})
  \rho_{i_1i_2} \otimes \cdots
  \otimes \rho_{i_{2N-1}i_{2N}} \;,
\end{equation}
where
\begin{equation}
p(i_1i_2\ldots i_{2N}) = \int d\w\, p(\w) p(i_1i_2\cdots i_{2N}|\w) \;.
\end{equation}

We now select a random subset of the bits $i_1\ldots i_{2N}$ and list all
the qubit pairs which have at least one associated bit in the subset.  From
this list we choose one qubit pair to be the {\it target}.  For each of
the other qubit pairs in the list, Alice and Bob both perform one of a set of
three unitary transformations on their half of the pair, followed by a
bilateral controlled-NOT onto the target pair.  This sequence of operations
is equivalent to replacing one of the bits of the target pair with the parity
of a subset of all the bits.  The choice of unitary transformation corresponds
to including the first, second, or both of the bits from a particular pair in
the parity calculation.  Then a measurement is performed on the target pair.
(The details of this procedure are given in \cite{Bennett1996c}.)
By carrying out such a procedure, one bit of joint information is acquired
about all the pairs, at the expense of sacrificing one entangled pair (that
is, two bits).  The unmeasured pairs in general undergo an invertible
transformation among the Bell states, but they do not become entangled with
each other, and this transformation can, if one chooses, be undone, leaving
the sequence of bits for the unmeasured pairs unaltered.  Bennett {\it et al.}\
have shown that such a procedure can be equivalent to finding the parity of
any subset of the $2N$ bits.  This parity bit then allows one to update the
probability distribution for the remaining $2(N-1)$-bit string.

Let us examine this in a little more detail.  Let
$\ivec \equiv i_1 i_2 \cdots i_{2N}$ denote a sequence of bits.  We
can select a subset of these bits by giving another sequence ${\vec x}$,
which includes a 1 for each bit to be included in the subset and a 0 for
the rest.  The parity of the subset is then
\begin{equation}
\pi_{\vec x}(\ivec) \equiv {\vec x} \cdot \ivec
  \equiv \Biggl( \sum_{m=1}^{2N} x_m i_m \Biggr) \bmod 2 \;.
\label{parity}
\end{equation}
For a given $\ivec$ the probability of getting a value
$\pi_{\vec x}$ for the parity is either 0 or 1, so the probability of
getting $\pi_{\vec x}$ as a measurement result is
\begin{equation}
p(\pi_{\vec x}) = \sum_\ivec p(\pi_{\vec x}|\ivec)
p(\ivec)
= \sum_\ivec \delta_{\pi_{\vec x},{\vec x}\cdot\ivec}
p(\ivec)\;.
\end{equation}
For simplicity let us assume that the target pair is the last, so the
last two bits are sacrificed; the new state for the $N-1$ remaining pairs is
\begin{equation}
\rho^{(N-1)} = \sum_{\ivec'} p({\ivec'}|\pi_{\vec x})
  \rho_{i_1i_2} \otimes \rho_{i_3i_4} \otimes \cdots \rho_{i_{2N-3}i_{2N-2}}\;,
\end{equation}
where $\ivec' \equiv i_1 i_2 \cdots i_{2N-2}$ and
\begin{equation}
p(\ivec'|\pi_{\vec x}) =
\sum_{i_{2N-1},i_{2N}}
p(\ivec|\pi_{\vec x}) =
\sum_{i_{2N-1},i_{2N}}
  {{ p(\ivec) p(\pi_{\vec x}|\ivec) }\over{ p(\pi_{\vec x}) }}
\;.
\end{equation}
Note that while the initial probability distribution $p(\ivec)$
is symmetric under interchanges of the pairs, this symmetry is lost after
measurement.

The purification scheme follows simply from this.  One chooses
subsets of the bit string at random and measures their parity,
sacrificing one pair with each measurement, but updating the
probability distribution for the remaining strings.  This
procedure is repeated until one is left with only a single string,
say $\ivec_0$, with probability $1-\delta$ for some small
$\delta$.  Written more formally, the posterior probability
$p_{\rm post}$ at the end of the procedure, conditioned on all
measurement results, has the property $p_{\rm
post}(\ivec_0)=1-\delta$ for some sequence $\ivec_0$.  One then
knows with high probability the maximally entangled state of each
remaining pair, which can then be transformed into a standard
state (such as $\psim$) by local operations.  The yield of this
procedure is the number of entangled pairs left at the end.

It is clear that there are states for which the yield is zero. The
obvious example is a state $\rho^{\otimes N}$ where $\rho$ is
unentangled. For states of the form $\rho_\w^{\otimes N}$, Bennett {\it et
al.}\ have shown that asymptotically the method gives a yield of $N(1-S_\w)$
maximally entangled pairs with fidelity approaching 1, where
\begin{equation}
S_\w = -\sum_{j=1}^4 w_j \log w_j = -\tr (\rho_\w \log\rho_\w)
\end{equation}
is the entropy of $\rho_\w$. The argument makes use of the theorem of typical
sequences \cite{Cover1991b} (which is closely related to Shannon's noiseless
coding theorem \cite{Shannon1948}), according to which, for any $\epsilon>0$
and $\delta>0$ and sufficiently large $N$, there exists a subset
${\rm TYP}(N)$ of the set of all sequences $\ivec$ with the following
properties:
\begin{equation}
p({\rm TYP}(N))\equiv
\sum_{\ivec\in {\rm TYP}(N)} p(\ivec|\vec w) \ge 1-\epsilon\;,
\end{equation}
i.e., the total probability of the set ${\rm TYP}(N)$ is arbitrarily
close to 1; and
\begin{equation}
|{\rm TYP}(N)| \le 2^{N(S_\w+\delta)} \;,
\end{equation}
i.e., the number of sequences in ${\rm TYP}(N)$ is not much larger than
$2^{NS_\w}$. The set ${\rm TYP}(N)$ is called the set of typical sequences.
Since the parity measurement in each hashing round rules out half the typical
sequences on average and since essentially all the probability is
concentrated on the typical sequences, it can be expected that after
sacrificing approximately $NS_\w$ pairs, essentially all the probability is
concentrated on a single typical sequence. Clearly this leads to a
positive yield only if $S_\w<1$.

The theorem of typical sequences does not hold in general for sequences
corresponding to exchangeable states of the form (\ref{rhon}). To apply the
hashing method in this case, we rely on a generalization of the theorem of
typical sequences due to Czisz\'ar and K\"orner \cite{Cziszar1981} (this
theorem has recently been used by Jozsa {\it et al.}\ \cite{Jozsa1998b} to
derive a universal quantum information compressing scheme). Applied to our
setting, the theorem is that, given a fixed entropy $S_0$, then for any
$\epsilon>0$ and $\delta>0$ and sufficiently large $N$, there exists a subset
${\rm CK}(N)$ of the set of all sequences $\ivec$ with the following
properties:
\begin{equation}
\sum_{\ivec\in {\rm CK}(N)} p(\ivec|\vec w)
\ge 1-\epsilon
\label{eq:CK1}
\end{equation}
for all $\w$ such that $S_\w<S_0$, which means that the set ${\rm CK}(N)$ is
typical for all probability distributions with entropy less than
$S_0$; and
\begin{equation}
|\,{\rm CK}(N)| \le 2^{N(S_0+\delta)} \;,
\label{eq:CK2}
\end{equation}
i.e., the number of sequences in ${\rm CK}(N)$ is not much larger than
$2^{NS_0}$.  In the following, when we write ``typical sequences,'' we
mean sequences in ${\rm CK}(N)$, whereas by ``atypical sequences'' we
mean sequences in $\overline{\rm CK}(N)$,  the complement of ${\rm CK}(N)$.

Now assume that we want to perform the hashing protocol on a state of
$N$ pairs of the form (\ref{rhon}) with the property
\begin{equation}
\int_{S_\w>S_0}d\w\,p(\w) = \eta \ll 1
\label{eq:defeta}
\end{equation}
for some entropy $S_0<1$; i.e., there is only a small {\it a priori\/}
probability that the entropy of the unknown state exceeds the given value
$S_0$.  (The case of states that do not have this property will be discussed
at the end of this section.)  Furthermore, assume that $N$ is large enough that
there exists a Czisz\'ar-K\"orner set ${\rm CK}(N)$ with constants
$\epsilon, \delta \ll 1$ in Eqs.~(\ref{eq:CK1}) and (\ref{eq:CK2}).
It then follows that
\begin{eqnarray}
p({\rm CK}(N))
&\equiv&\sum_{\ivec\in {\rm CK}(N)} p(\ivec) \cr
&=&  \sum_{\ivec\in {\rm CK}(N)} \int d\w\, p(\w) p(\ivec|\w) \cr
&\ge& \sum_{\ivec\in {\rm CK}(N)} \int_{S_\w<S_0} d\w\,
p(\w) p(\ivec|\w) \cr
&=&  \int_{S_\w<S_0} d\w\, p(\w)\sum_{\ivec\in {\rm CK}(N)}
p(\ivec|\w) \cr
&\ge&  \int_{S_\w<S_0} d\w\, p(\w) (1-\epsilon) \cr
&=& \vphantom{\int_{S_\w<S_0}} (1-\eta)(1-\epsilon) \cr
&\ge&  \vphantom{\int_{S_\w<S_0}} 1-\eta-\epsilon \;,
\label{eq:CK3}
\end{eqnarray}
where Eqs.~(\ref{rhon_norm}), (\ref{eq:CK1}), and (\ref{eq:defeta}) have been
used.

We use this inequality, in combination with Eq.~(\ref{eq:CK2}), to
derive the asymptotic yield of the hashing algorithm applied to
exchangeable states. We restrict our analysis to a simplified
protocol, in which we choose a number $r$, somewhat larger than
$N(S_0+\delta)$, such that
\begin{equation}
\zeta \equiv 2^{N(S_0+\delta)-r} \ll 1 \;.
\end{equation}
We begin with input strings $\ivec$ that have probability
$p(\ivec)$. Let $h$ denote a sequence of $r$ parity checks on
random subsets, and let $\ovec=o_1,\ldots,o_r$ denote the $r$-bit
string of parity checks, or outcomes. Note that we denote all
strings of bits as vectors, even though they are not all of the
same length. The probability distribution $p(h)$ on parity-check
sequences is weighted uniformly on all sequences.  For a given
input string $\ivec$ and a given parity-check sequence $h$, the
outcome $\ovec$ is determined; we denote this deterministic
outcome by $\ovec_{h,\ivec}$. We can express this deterministic
outcome in terms of a probability for outcome string $\ovec$,
given parity-check sequence $h$ and input string $\ivec$:
\begin{equation}
p(\ovec|h,\ivec)=\delta_{\ovec,\ovec_{h,\ivec}}\;.
\label{eq:pohi}
\end{equation}

Since for each parity-check bit obtained, two bits of the input
string are discarded, two strings with the same parity check,
which differ only on those two bits, become the same after that
step. After $r$ steps of a parity-check sequence $h$, there will
be only $N-r$ entangled pairs, corresponding to a string of
$2(N-r)$ bits. If one starts with a string $\ivec$, one will be
left with a shorter substring $\ivec_h(\ivec)$.  Different initial
strings $\ivec$ that generate the same outcome $\ovec$ and lead to
the same final substring $\ivec_h(\ivec)$ are equivalent for
practical purposes.  Let us denote the set of all {\it input\/}
strings $\ivec$ which lead to outcome $\ovec$ and to {\it output\/}
substring $\ivec_h$ by $I_h\!(\ovec,\ivec_h) \equiv
\{\ivec|\ovec_{h,\ivec}=\ovec,\ivec_h(\ivec)=\ivec_h\}$.

For parity-check sequence $h$, we are interested in outcomes
$\ovec$ such that all {\it typical\/} input strings $\ivec$ that
lead to $\ovec$ produce the same output string $\ivec_h(\ivec)$.
For outcomes where this is the case, the procedure picks out a
unique output string from among all those that could be produced
by a typical input string.  In this case, we say that we {\it
accept\/} the outcome $\ovec$ and the corresponding unique output
string, which we denote by $\ivec_{h,\ovec}$.  In this way we
divide the outcomes for a parity-check sequence $h$ into two sets,
the set of accepted outcomes, $A_h$, and its complement.  For an
outcome that we accept and for a typical input string, we can
write the conditional probability~(\ref{eq:pohi}) as
\begin{eqnarray}
p(\ovec|h,\ivec,\ivec\in\CK(N))
&=&\cases{
    1\;,&if $\ivec\in I_h(\ovec,\ivec_{h,\ovec})$\cr
    0\;,&if $\ivec\notin I_h(\ovec,\ivec_{h,\ovec})$}
\nonumber\\
&=&\delta_{\ovec,\ovec_{h,\ivec}}\delta_{\ivec_h(\ivec),\ivec_{h,\ovec}}
\quad\mbox{for $\ovec\in A_h$.}
\label{eq:condoih}
\end{eqnarray}
Though the additional Kronecker delta in this expression is
redundant, it reminds one that any {\it typical\/} input string
$\ivec$ that leads to an {\it accepted\/} outcome $\ovec$ produces
output string $\ivec_{h,\ovec}$.  Notice that this is not true for
atypical input strings: an atypical input string can have outcome
$\ovec$ and produce outcome string $\ivec_{h,\ovec}$ or a
different output string.

The probability that the outcome is accepted, given input
string $\ivec$ and parity-check sequence $h$, is
\begin{eqnarray}
p({\rm accept}|h,\ivec)
&=&\sum_{\ovec\in A_h}p(\ovec|h,\ivec)\nonumber\\
&=&\sum_{\ovec\in A_h}\delta_{\ovec,\ovec_{h,\ivec}}\nonumber\\
&=&\cases{
        1\;,&if $\ivec$ leads to an accepted outcome,\cr
        0\;,&if $\ivec$ does not lead to an accepted outcome.
         }
\end{eqnarray}
Notice that this conditional acceptance probability can be nonzero
for atypical input strings.  The complementary probability, that the outcome
is not accepted, given $\ivec$ and $h$, is given by
\begin{eqnarray}
p(\overline{\rm accept}|h,\ivec)
&=&\sum_{\ovec\notin A_h}p(\ovec|h,\ivec)\nonumber\\
&=&\sum_{\ovec\notin A_h}\delta_{\ovec,\ovec_{h,\ivec}}\nonumber\\
&=&\cases{
        0\;,&if $\ivec$ leads to an accepted outcome,\cr
        1\;,&if $\ivec$ does not lead to an accepted outcome.
         }
\end{eqnarray}
If the input string is a typical string, the conditional
acceptance probability can also be written as
\begin{equation}
p({\rm accept}|h,\ivec,\ivec\in\CK(N))
=\sum_{\ovec\in A_h}\delta_{\ovec,\ovec_{h,\ivec}}
\delta_{\ivec_h(\ivec),\ivec_{h,\ovec}}
\end{equation}
[see Eq.~(\ref{eq:condoih})].

What we are interested in for the present is the probability to
have an outcome that is accepted, given a typical input string, but
averaged over all parity-check sequences:
\begin{eqnarray}
p({\rm accept}|\ivec,\ivec\in\CK(N))
&=&\sum_h p({\rm accept}|h,\ivec,\ivec\in\CK(N))p(h)\nonumber\\
&=&\sum_h p(h)\sum_{\ovec\in A_h}\delta_{\ovec,\ovec_{h,\ivec}}
\delta_{\ivec_h(\ivec),\ivec_{h,\ovec}}\;.
\label{eq:paccepti}
\end{eqnarray}
The complementary probability,
\begin{eqnarray}
p(\overline{\rm accept}|\ivec,\ivec\in\CK(N))
&=&\sum_h p(\overline{\rm accept}|h,\ivec,\ivec\in\CK(N))p(h)\nonumber\\
&=&\sum_h p(h)\sum_{\ovec\notin A_h}\delta_{\ovec,\ovec_{h,\ivec}}\;,
\end{eqnarray}
is the average probability not to have an outcome that is
accepted, given the typical input string $\ivec$.  This
probability is the probability that for a random parity-check
sequence, the typical input string $\ivec$ leads to an outcome
that does not pick out a unique output string $\ivec_{h,\ovec}$,
i.e., does not lead to the only possible output string that could
have been produced by a typical input string.  We can bound this
probability in the following way.  The number of typical sequences
satisfies $|\CK(N)|\le 2^{N(S_0+\delta)}$.  For parity subsets
chosen randomly, the probability that two typical input strings,
$\ivec$ and $\j$, agree on all $r$ parity checks---i.e., have the
same outcome---is $\le2^{-r}$; thus the probability that $\ivec$
and $\j$ agree on all $r$ parity checks {\it and\/} produce {\it
different\/} output strings, $\ivec_h(\ivec)$ and $\ivec_h(\j)$,
is $\le2^{-r}$.  Hence the probability of not producing a unique
output, given a typical input $\ivec$, is bounded by
\begin{equation}
p(\overline{\rm accept}|\ivec,\ivec\in\CK(N))
\le2^{-r}\times2^{N(S_0+\delta)} = \zeta\;.
\end{equation}
This implies that the conditional acceptance probability~(\ref{eq:paccepti})
satisfies
\begin{equation}
p({\rm accept}|\ivec,\ivec\in\CK(N))\ge1-\zeta\;.
\label{eq:pacceptibound}
\end{equation}

Bayes's rule tells us that the posterior probability for output string
$\ivec_h$, given $h$ and $\ovec$, is
\begin{eqnarray}
p(\ivec_h|h,\ovec)&=&
\sum_{\ivec\in I_h\!(\ovec,\ivec_h)}p(\ivec|h,\ovec)\nonumber\\
&=&\sum_{\ivec\in I_h\!(\ovec,\ivec_h)}
   {p(\ovec|h,\ivec)p(h)p(\ivec)\over p(\ovec|h)p(h)}\nonumber\\
&=&\sum_{\ivec\in I_h\!(\ovec,\ivec_h)}
   {p(\ovec|h,\ivec)p(\ivec)\over p(\ovec|h)}\;,
\end{eqnarray}
where
\begin{equation}
p(\ovec|h)=\sum_\ivec p(\ovec|h,\ivec)p(\ivec)
\end{equation}
is the probability for outcome string $\ovec$, given parity-check sequence
$h$.

Given a parity-check sequence $h$ and an accepted outcome $\ovec\in A_h$
for that sequence, we judge the ``success'' of the accepted
output string $\ivec_{h,\ovec}$ by the posterior probability, i.e,
\begin{equation}
p({\rm success}|h,\ovec)=
p(\ivec_{h,\ovec}|h,\ovec)
=\sum_{\ivec\in I_h\!(\ovec,\ivec_{h,\ovec})}p(\ivec|h,\ovec)
\quad\mbox{for $\ovec\in A_h$.}
\end{equation}
The total probability of success, $p({\rm success})$, is obtained
by averaging over all parity-check sequences $h$ and over all
accepted outcomes $\ovec\in A_h$.  This probability can be manipulated
in the following ways:
\begin{eqnarray}
p({\rm success})
&=&\sum_h\sum_{\ovec\in A_h}p({\rm success}|h,\ovec)p(\ovec|h)p(h)\nonumber\\
&=&\sum_h\sum_{\ovec\in A_h}\sum_{\ivec\in I_h\!(\ovec,\ivec_{h,\ovec})}
p(\ivec|h,\ovec)p(\ovec|h)p(h)\nonumber\\
&=&\sum_h\sum_{\ovec\in A_h}\sum_{\ivec\in I_h\!(\ovec,\ivec_{h,\ovec})}
p(\ovec|h,\ivec)p(h)p(\ivec)\nonumber\\
&\ge&\sum_h\sum_{\ovec\in A_h}
\sum_{{\scriptstyle{\ivec\in I_h\!(\ovec,\ivec_{h,\ovec})}}
\atop{\scriptstyle{\ivec\in\CK(N)}}}
p(\ovec|h,\ivec)p(h)p(\ivec)\nonumber\\
&=&\sum_h\sum_{\ovec\in A_h}\sum_{\ivec\in\CK(N)}
  \delta_{\ovec,\ovec_{h,\ivec}}
  \delta_{\ivec_h(\ivec),\ivec_{h,\ovec}}p(h)p(\ivec)\;.
\end{eqnarray}
The inequality here follows from restricting the sum over input
strings to typical strings and reflects the fact that an atypical
string might lead to an accepted outcome {\it and\/} to the accepted
output string $\ivec_{h,\ovec}$, thereby contributing to the
success probability.  The final equality comes from using
Eq.~(\ref{eq:condoih}) for $p(\ovec|h,\ivec)$.  Using
Eqs.~(\ref{eq:CK3}), (\ref{eq:paccepti}), and
(\ref{eq:pacceptibound}), we can now bound the probability of
success:
\begin{eqnarray}
p({\rm success})
&\ge&\sum_{\ivec\in\CK(N)}p(\ivec)\sum_h p(h)\sum_{\ovec\in A_h}
  \delta_{\ovec,\ovec_{h,\ivec}}
  \delta_{\ivec_h(\ivec),\ivec_{h,\ovec}}\nonumber\\
&=&\sum_{\ivec\in\CK(N)}p(\ivec)p({\rm accept}|\ivec,\ivec\in{\rm CK}(N))
    \nonumber\\
&\ge&(1-\zeta)\sum_{\ivec\in\CK(N)}p(\ivec)\nonumber\\
&=&\vphantom{\sum_{\ivec}}(1-\zeta)p(\CK(N))\nonumber\\
&\ge&\vphantom{\sum_{\ivec}}(1-\zeta)(1-\eta-\epsilon)\nonumber\\
&\ge&\vphantom{\sum_{\ivec}}1-\zeta-\eta-\epsilon\;.
\label{bound}
\end{eqnarray}
This is the desired result. Assuming we can choose arbitrary
positive constants $\epsilon$ and $\eta$ and have sufficiently
large $N$, the probability~(\ref{bound}) can be made arbitrarily
close to 1.

Except for certain singular distributions $p(\w)$, given an
exchangeable state of the form~(\ref{rhon}), it is always possible
to make $\eta$ in Eq.~(\ref{eq:defeta}) arbitrarily small by
choosing the entropy $S_0$ sufficiently large ($0\le S_0<2$); if
$S_0\ge1$, however, then the number of hashing rounds $r\ge N$,
which means there is no yield since $N-r\le0$. To decrease the
value of $S_0$ and thereby make the yield positive or increase an
already positive yield, one can perform quantum state tomography
on some of the pairs to obtain more data about the state,
generally producing a narrower posterior distribution $p'(\w)$
(see Sec.~\ref{sec:assign}). The width of the posterior
distribution depends on the number of pairs sacrificed for the
tomographic measurements, but not on the total number of pairs
$N$. The number of pairs needed for tomography can therefore be
neglected in the asymptotic limit of large $N$.

Asymptotically, the prior probability of obtaining a posterior
$p'(\w)$ concentrated at $\w=\w_0$ with an entropy $S_{\w_0}<S_0$
is given by the expression
\begin{equation}
p(S<S_0) \equiv \int_{S_\w<S_0}d\w\,p(\w) \;,
\end{equation}
where $p(\w)$ is the prior distribution~(\ref{rhon}) defining the
initial state. Putting everything together we see that, for
$S_0<1$, $p(S<S_0)$ is the probability of obtaining an asymptotic
yield of $N(1-S_0)$ using a combination of quantum state
tomography and one-way hashing.

If most of the prior distribution $p(\w)$ is concentrated on states with an
entropy exceeding 1 bit, i.e., if $p(S<1)$ is small, then it will
normally be a better strategy to precede the hashing procedure by a few
iterations of the recurrence method. This is the content of the next section.

\section{Entanglement purification by recurrence}  \label{sec:recurrence}

If the generating function $p(\w)$ has no significant support on weights $\w$
with $S_\w < 1$, then hashing cannot be used for entanglement purification,
at least initially.  It might still be possible, however, to distill some
entanglement by using the more robust (but far more wasteful) technique
of {\it recurrence\/} \cite{Bennett1996c,Deutsch1998}.

In the recurrence algorithm, an initial set of $2N$ entangled qubit pairs is
grouped into $N$ sets of 2 pairs each.  In each set, one pair is designated
the {\it target\/} pair, and the other the {\it control\/} pair. Alice and Bob
thus have $N$ target qubits and $N$ control qubits each. Alice now rotates all
her qubits by $\pi/2$ about the $x$ axis, while Bob rotates all his qubits by
$-\pi/2$ about the $x$ axis.  Each of them then performs a controlled-NOT operation
from each control qubit onto the corresponding target qubit and measures his
or her target qubit in the $z$ basis ($|0\rangle$ and $|1\rangle$).  The
target qubits are then discarded.  If Alice and Bob both get the same result
for a given target pair (i.e., both 0 or both 1), the procedure has succeeded,
and the control pair can be shown to have increased entanglement.  If their
results differ, the procedure has failed, and the control qubits must also be
discarded.

If the state of both target and control pairs is of form (\ref{bell_mixtures}),
the probability of success is
\begin{equation}
\psucc = \psucc(\w) =  (w_1+w_4)^2 + (w_2+w_3)^2\;,
\end{equation}
and the new state of the control pair after the measurement has weights
\cite{Deutsch1998}
\begin{eqnarray}
w_1' &=& 2 w_2 w_3/\psucc\;, \nonumber\\
w_2' &=& (w_2^2+w_3^2)/\psucc\;, \nonumber\\
w_3' &=& 2 w_1 w_4/\psucc\;, \nonumber\\
w_4' &=& (w_1^2+w_4^2)/\psucc \;.
\label{neww}
\end{eqnarray}
If initially $w_4 > 1/2$, then this procedure converges towards $w_4=1$.
The convergence is slow, however, and since more than half of all the pairs
is discarded each time, the yield is generally low.

Suppose that instead of a product state we have an exchangeable state
of the form (\ref{rhon}).  We can carry out the procedure exactly as
before, grouping the pairs into sets of two, with a target and control
bit.  If there are initially $2N$ pairs in the state
\begin{equation}
\rho^{(2N)} = \int d\w\, p(\w) \rho_\w^{\otimes 2N} \;,
\label{rho2n}
\end{equation}
then after performing the measurements, Alice and Bob will get the
same result $\Nsucc$ times and different results $N-\Nsucc$ times,
leaving them with a new state of the form (\ref{rho2n}) for
$\Nsucc$ pairs. For large $N$, the posterior distribution
$p(\w|\Nsucc)$ will generally be sharply peaked about those $\w$
which give a value of $\psucc$ close to $\Nsucc/N$.  Unlike
hashing, the recurrence algorithm produces a posterior state
$\rho^{(\Nsucc)}$ which is exchangeable.  We now turn to how we
find this state in light of the measurement results.

Compared with the hashing algorithm, where precisely one bit of information
is obtained in each round of the procedure, in the recurrence method
much more information is obtained, namely the value of $\Nsucc$.
We can therefore deduce the posterior distribution
\begin{equation}
p(\w|\Nsucc) =   {p(\Nsucc|\w) p(\w) \over p(\Nsucc)} \;,
\end{equation}
where
\begin{equation}
p(\Nsucc|\w) = {N\choose\Nsucc} [\psucc(\w)]^\Nsucc
  [1-\psucc(\w)]^{N-\Nsucc}\;,
\end{equation}
and
\begin{equation}
p(\Nsucc) = \int d\w\, p(\Nsucc|\w) p(\w) \;.
\end{equation}
Because the remaining states have been transformed according to
(\ref{neww}), we must also change to the new variables $\w^{\,\prime}$.  So the
new state is
\begin{equation}
{\tilde\rho}^{(\Nsucc)} =
\int d\w^{\,\prime}\, p'(\w^{\,\prime}) \rho(\w^{\,\prime})\;,
\end{equation}
where
\begin{equation}
p'(\w^{\,\prime})\,d\w^{\,\prime} = p(\w|\Nsucc)\,d\w\;.
\end{equation}

While this Bayesian procedure is very simple compared to the hashing
method, it is still a bit too complicated for simple illustration.  There is,
however, an even simpler variant of this technique that is easy to analyze.
Suppose that, instead of the general Bell-diagonal state (\ref{bell_mixtures}),
we have an initial Werner state
\begin{equation}
\rho(F) = F \phip + {1-F\over3}(\phim+\psip+\psim)\;.
\label{werner}
\end{equation}
We can carry out the recurrence procedure exactly as above, with the
probability of success
\begin{equation}
\psucc(F) = (8 F^2 - 4 F + 5)/9 \;;
\end{equation}
here $F$ denotes the fidelity of the state with $\phip$, with $F>1/2$
necessary for distillability.  The recurrence procedure
does not in general lead to a new state of form (\ref{werner}),
but by twirling the state can be put in this form, at the
cost of some increase in entropy.  The new state has a fidelity
\begin{equation}
F' = { {10 F^2 - 2 F + 1} \over {8 F^2 - 4 F + 5} } \;.
\label{new_fidelity}
\end{equation}

Suppose that we have $2N$ entangled pairs, with partial information
sufficient to determine that they are all in a state of the form~(\ref{werner}),
but not to determine the exact fidelity $F$.  The joint
state of the pairs is then
\begin{equation}
\rho^{(2N)} = \int dF\, p(F) \rho(F)^{\otimes 2N} \;.
\end{equation}
We then group the pairs into sets of two and carry out the recurrence
procedure on each set, with $\Nsucc$ successful results.  We can then
deduce a revised generating function
\begin{equation}
p(F|\Nsucc) =  { p(\Nsucc|F) p(F) \over p(\Nsucc)}\;,
\end{equation}
where
\begin{equation}
p(\Nsucc|F) = {N\choose\Nsucc} [\psucc(F)]^\Nsucc
  [1-\psucc(F)]^{N-\Nsucc}\;,
\end{equation}
and
\begin{equation}
p(\Nsucc) = \int dF\, p(\Nsucc|F) p(F) \;.
\end{equation}
The new density operator for the $\Nsucc$ remaining pairs is
\begin{equation}
\rho^{(\Nsucc)} = \int dF'\, p'(F') \rho(F')^{\otimes \Nsucc} \;,
\end{equation}
where the the posterior distribution is expressed in terms of the new
variable $F'$ given by (\ref{new_fidelity}).  Working this out explicitly,
we get
\begin{equation}
p'(F') = \left(8 F(F') - 2
  + {3(3-4F')\over\sqrt{6F'-4{F'}^2-1}}\right)
  { { p[F(F')|\Nsucc] } \over 10-8F' } \;,
\end{equation}
where $F(F')$ is the inverse of (\ref{new_fidelity}):
\begin{equation}
F(F') = { { (1 - 2F') + 3 \sqrt{6F' - 4{F'}^2 - 1} }
  \over { 10 - 8 F' } } \;.
\end{equation}

We can see how much information is gained by a single round of the recurrence
method using this simplified version as an example.  If the initial generating
function is a uniform distribution, $p(F) = 4/3$ for $1/4<F<1$, then for large
$N$ the posterior distribution is highly peaked after one round.  We see this
in Figure 1, where the prior and posterior distributions are shown for
different values of $N$ and a typical choice of $\Nsucc$.  Note that states
with $1/4<F<1/2$ move towards $F=1/4$ under the procedure, producing a
peak about the completely mixed state; for high $N$ and the value of $\Nsucc$
used in our example, this peak is suppressed by the Bayesian updating.  States
with $F>1/2$ move towards $F=1$.  The procedure has fixed points at $F=1/4$,
$F=1/2$, and $F=1$.

It should be noted that because of its extremely small yield, the recurrence
method should never be used if hashing is possible.  An initial state that
cannot be distilled by the hashing method, however, might, after one or more
rounds of the recurrence method, satisfy the criterion (\ref{eq:defeta}) for
some value of $S_0<1$.  If that is so, then a combination of tomography and
hashing should be used thereafter, as described in the last section.

Similarly, if $p(\rho)$ has some support on distillable and some on
undistillable states, a few rounds of the recurrence method
generally produces convergence on either a distillable or undistillable
state, without ambiguity.  Under certain circumstances, however, it
might be beneficial to supplement this with tomographic measurements on a number
of pairs as well. For example, the updating procedure
(\ref{neww}) treats the coefficients $w_1,w_4$ and $w_2,w_3$
symmetrically.  An initially symmetric state thus has this
symmetry preserved, and the distribution $p(\w)$ might become double-peaked.
In this case, measuring a small number of pairs would suffice
to eliminate one of the two peaks.

\section{Conclusion}       \label{sec:conclude}

We have given a Bayesian account of the entanglement purification procedures
of one-way hashing and recurrence. The Bayesian formulation allows us to
provide a straightforward discussion of the conditions under which maximally
entangled states can be distilled from unknown or partially known quantum
states. For one-way hashing, we have given the {\it a priori\/} probabilities
for the possible asymptotic yields of maximally entangled pairs. Our results
can be used to decide which combination of quantum state tomography,
recurrence, and hashing to use to obtain the highest expected yield,
both asymptotically and in the case of a fixed number of initially given
pairs. Although our discussion is entirely in terms of pairs of qubits, the
method is general and can be applied to any generalization of
hashing or recurrence in Hilbert spaces of higher dimension.

\section*{Acknowledgments}

We would like to thank Howard Barnum, Oliver Cohen, Chris Fuchs, and
Bob Griffiths for helpful conversations.  T.A.B. was supported in part by
NSF Grant No.~PHY-9900755 and DOE Grant No.~DE-FG02-90ER40542,
R.S. was supported by the UK Engineering and Physical Sciences
Research Council, and C.M.C. was supported in part by ONR Grant
No.~N00014-00-1-0578.  Some of this work was done at the July 2000 workshop
on ``Quantum Information Processing'' at the Benasque Center for Science
in Benasque, Spain.


\begin{thebibliography}{10}

\bibitem{Bennett1996b}
C.~H. Bennett {\it et~al.}, Phys.\ Rev.\ Lett.\ {\bf 76},  722  (1996).

\bibitem{Bennett1996c}
C.~H. Bennett, D.~P. DiVincenzo, J.~A. Smolin, and W.~K. Wootters, Phys.\ Rev.\
  A {\bf 54},  3824  (1996).

\bibitem{Horodecki1997b}
M. Horodecki, P. Horodecki, and R. Horodecki, Phys.\ Rev.\ Lett.\ {\bf 78},
  574  (1997).

\bibitem{Deutsch1998}
D. Deutsch {\it et~al.}, Phys.\ Rev.\ Lett.\ {\bf 80},  2022  (1998).

\bibitem{Giedke1999}
G. Giedke, H.-J. Briegel, J.~I. Cirac, and P. Zoller, Phys.\ Rev.\ A {\bf 59},
  2641  (1999).

\bibitem{Eisert2000a}
J. Eisert {\it et~al.}, Phys.\ Rev.\ Lett.\ {\bf 84},  1611  (2000).

\bibitem{Horodecki1999b}
R. Horodecki, M. Horodecki, and P. Horodecki, Phys.\ Rev.\ A {\bf 59},  1799
  (1999).

\bibitem{Jaynes1957a}
E.~T. Jaynes, Phys.\ Rev.\ {\bf 106},  620  (1957).

\bibitem{Jaynes1957b}
E.~T. Jaynes, Phys.\ Rev.\ {\bf 108},  171  (1957).

\bibitem{Rajagopal1999}
A.~K. Rajagopal, Phys.\ Rev.\ A {\bf 60},  4338  (1999).

\bibitem{Rigo2000}
A. Rigo, A.~R. Plastino, A. Plastino, and M. Casas, Phys.\ Lett.\ A {\bf 270},
  1  (2000).

\bibitem{Linden1998b}
N. Linden, S. Massar, and S. Popescu, Phys.\ Rev.\ Lett.\ {\bf 81},  3279
  (1998).

\bibitem{Jaynes1986}
E.~T. Jaynes,  in {\em Maximum Entropy and Bayesian Methods in Applied
  Statistics}, edited by J.~H. Justice (Cambridge University Press, Cambridge,
  1986), pp.\ 26--58.

\bibitem{DeFinetti1990}
B. de~Finetti, {\em Theory of Probability\/} (Wiley, New York, 1990).

\bibitem{Hudson1976}
R.~L. Hudson and G.~R. Moody, Z. Wahrscheinlichkeitstheorie verw.\ Geb.\ {\bf
  33},  343  (1976).

\bibitem{Stormer1969}
E. St{\o}rmer, Journal of Functional Analysis {\bf 3},  48  (1969).

\bibitem{Caves2000a}
C.~M. Caves, C.~A. Fuchs, and R. Schack, to be published.

\bibitem{Bures1969}
D. Bures, Trans.\ Am.\ Math.\ Soc.\ {\bf 135},  199  (1969).

\bibitem{Braunstein1994a}
S.~L. Braunstein and C.~M. Caves, Phys.\ Rev.\ Lett.\ {\bf 72},  3439  (1994).

\bibitem{Slater1997}
P.~B. Slater, J. Math.\ Phys.\ {\bf 38},  2274  (1997).

\bibitem{Zyczkowski1998b}
K. \.Zyczkowski, P. Horodecki, A. Sanpera, and M. Lewenstein, Phys.\ Rev.\ A
  {\bf 58},  883  (1998).

\bibitem{Skilling1989a}
One possibility, whose analogue for classical probabilities was proposed by
Skilling [J.~Skilling, in {\em Maximum Entropy and {B}ayesian Methods},
edited by J.~Skilling (Kluwer, Dordrecht, 1989), pp.~45--52] is to use
$f(\rho)=e^{\alpha S(\rho)}$, along with one of the proposed unbiased
measures for $d\rho$ \cite{Bures1969,Braunstein1994a,Slater1997,Zyczkowski1998b}.
Here $\alpha\gg1$ is a parameter that characterizes one's confidence in
the single-copy maximum-entropy assignment $\rho_J$: for $N\ll\alpha$, the
exchangeable state~(\ref{eq:exch}) becomes effectively the product state
$\rho_J^{\otimes N}$, but for $N\gg\alpha$, Eq.~(\ref{eq:exch}) predicts
measurement statistics different from the product state.

\bibitem{Kraus1983}
K. Kraus, {\em States, Effects, and Operations. Fundamental Notions of Quantum
  Theory\/} (Springer, Berlin, 1983), lecture Notes in Physics Vol.\ 190.

\bibitem{Schack2000a}
R. Schack, T.~A. Brun, and C.~M. Caves, quant-ph/0008113.

\bibitem{Vogel1989b}
K. Vogel and H. Risken, Phys.\ Rev.\ A {\bf 40},  2847  (1989).

\bibitem{Smithey1993}
D.~T. Smithey, M. Beck, M.~G. Raymer, and A. Faridani, Phys.\ Rev.\ Lett.\
  {\bf 70},  1244  (1993).

\bibitem{Leonhardt1995}
U. Leonhardt, Phys.\ Rev.\ Lett.\ {\bf 74},  4101  (1995).

\bibitem{Cover1991b}
T.~M. Cover and J.~A. Thomas, {\em Elements of Information Theory\/} (Wiley, New
  York, 1991).

\bibitem{Shannon1948}
C.~E. Shannon, Bell Syst.\ Tech.\ J. {\bf 27},  379  (1948).

\bibitem{Cziszar1981}
I. Czisz\'ar and J. K\"orner, {\em Information Theory: Coding Theorems for
  Discrete Memoryless Systems\/} (Academic Press, New York, 1981).

\bibitem{Jozsa1998b}
R. Jozsa, M. Horodecki, P. Horodecki, and R. Horodecki, Phys.\ Rev.\ Lett.\
  {\bf 81},  1714  (1998).

\end{thebibliography}

\eject

\begin{figure}
\begin{center}
\epsfig{file=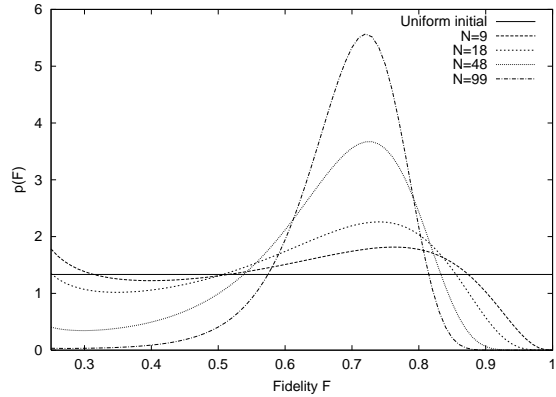, width=3in}
\label{fig1}
\end{center}
\caption{Plots of an initially uniform distribution for the generalized
Werner state for fidelities between $F=1/4$ (maximally mixed) and $F=1$
(maximally entangled) and updated distributions after one round of the
simplified recurrence method.  Before the round there are $2N$ pairs;
we assume the procedure succeeds in $2N/3$ cases ($\psucc=2/3$).  The new
distribution is plotted for $N=9,18,48,99$.  The new distribution is more
and more highly peaked for bigger $N$, and the probability of unentangled
states is more and more strongly suppressed.}
\end{figure}

\end{document}